\documentclass[sigconf]{acmart}
\usepackage{subfigure}
\usepackage{graphicx}
\usepackage{enumitem}
\usepackage{multirow}
\usepackage{makecell}
\usepackage{graphicx}
\usepackage{subfloat}
\usepackage{color}
\usepackage{caption}
\usepackage{subcaption}
\usepackage{enumitem}
\usepackage{xspace}
\usepackage{mdframed}
\usepackage{balance}

\AtBeginDocument{%
  }

\copyrightyear{2025}
\acmYear{2025}
\setcopyright{acmlicensed}
\acmConference[KDD '25]{Proceedings of the 31st ACM SIGKDD Conference on Knowledge Discovery and Data Mining V.1}{August 3--7, 2025}{Toronto, ON, Canada}
\acmBooktitle{Proceedings of the 31st ACM SIGKDD Conference on Knowledge Discovery and Data Mining V.1 (KDD '25), August 3--7, 2025, Toronto, ON, Canada}
\acmDOI{10.1145/3690624.3709440}
\acmISBN{979-8-4007-1245-6/25/08}
\newcommand{\method}{NoteLLM-2\xspace}

\author{Chao Zhang}
\affiliation{
 \institution{University of Science and Technology of China \& City University of Hong Kong \& State Key Laboratory of Cognitive Intelligence}
 \city{Hefei}
\country{China}
 }
\email{zclfe00@gmail.com}

 \author{Haoxin Zhang}
\affiliation{
 \institution{Xiaohongshu Inc.}
 \city{Beijing}
\country{China}
 }
\email{zhanghaoxin@xiaohongshu.com}

 \author{Shiwei Wu}
\affiliation{
 \institution{University of Science and Technology of China \& State Key Laboratory of Cognitive Intelligence}
 \city{Hefei}
\country{China}
 }
\email{dwustc@mail.ustc.edu.cn}

\author{Di Wu}
\affiliation{
 \institution{Xiaohongshu Inc.}
 \city{Beijing}
\country{China}
 }
\email{wudi1123@foxmail.com}

 \author{Tong Xu}
 \authornote{Corresponding authors.}
\affiliation{
 \institution{University of Science and Technology of China \& State Key Laboratory of Cognitive Intelligence}
 \city{Hefei}
\country{China}
 }
\email{tongxu@ustc.edu.cn}

 \author{Xiangyu Zhao}
 \authornotemark[1]
\affiliation{
 \institution{City University of Hong Kong}
 \city{Hong Kong}
\country{China}
 }
\email{xianzhao@cityu.edu.hk}

 \author{Yan Gao}
\affiliation{
 \institution{Xiaohongshu Inc.}
 \city{Beijing}
\country{China}
 }
\email{yadun@xiaohongshu.com}

 \author{Yao Hu}
\affiliation{
 \institution{Xiaohongshu Inc.}
 \city{Beijing}
\country{China}
 }
\email{xiahou@xiaohongshu.com}

 \author{Enhong Chen}
 \authornotemark[1]
\affiliation{
 \institution{University of Science and Technology of China \& State Key Laboratory of Cognitive Intelligence}
 \city{Hefei}
\country{China}
 }
\email{cheneh@ustc.edu.cn}
\renewcommand{\shortauthors}{Chao Zhang et al.}


\begin{document}

\title{\method: Multimodal Large Representation Models for Recommendation}

\begin{abstract}
Large Language Models (LLMs) have demonstrated exceptional proficiency in text understanding and embedding tasks. 
However, their potential in multimodal representation, particularly for item-to-item (I2I) recommendations, remains underexplored. 
While leveraging existing Multimodal Large Language Models (MLLMs) for such tasks is promising, challenges arise due to their delayed release compared to corresponding LLMs and the inefficiency in representation tasks. 
To address these issues, we propose an end-to-end fine-tuning method that customizes the integration of any existing LLMs and vision encoders for efficient multimodal representation. 
Preliminary experiments revealed that fine-tuned LLMs often neglect image content.
To counteract this, we propose \method, a novel framework that enhances visual information. 
Specifically, we propose two approaches: 
first, a prompt-based method that segregates visual and textual content, employing a multimodal In-Context Learning strategy to balance focus across modalities; 
second, a late fusion technique that directly integrates visual information into the final representations. 
Extensive experiments, both online and offline, demonstrate the effectiveness of our approach.
Code is available at \url{https://github.com/Applied-Machine-Learning-Lab/NoteLLM}.
\end{abstract}



\maketitle

\section{Introduction}
\label{para:sec1}
\begin{figure*}[htbp]
    \centering
    \includegraphics[width=0.95\textwidth]{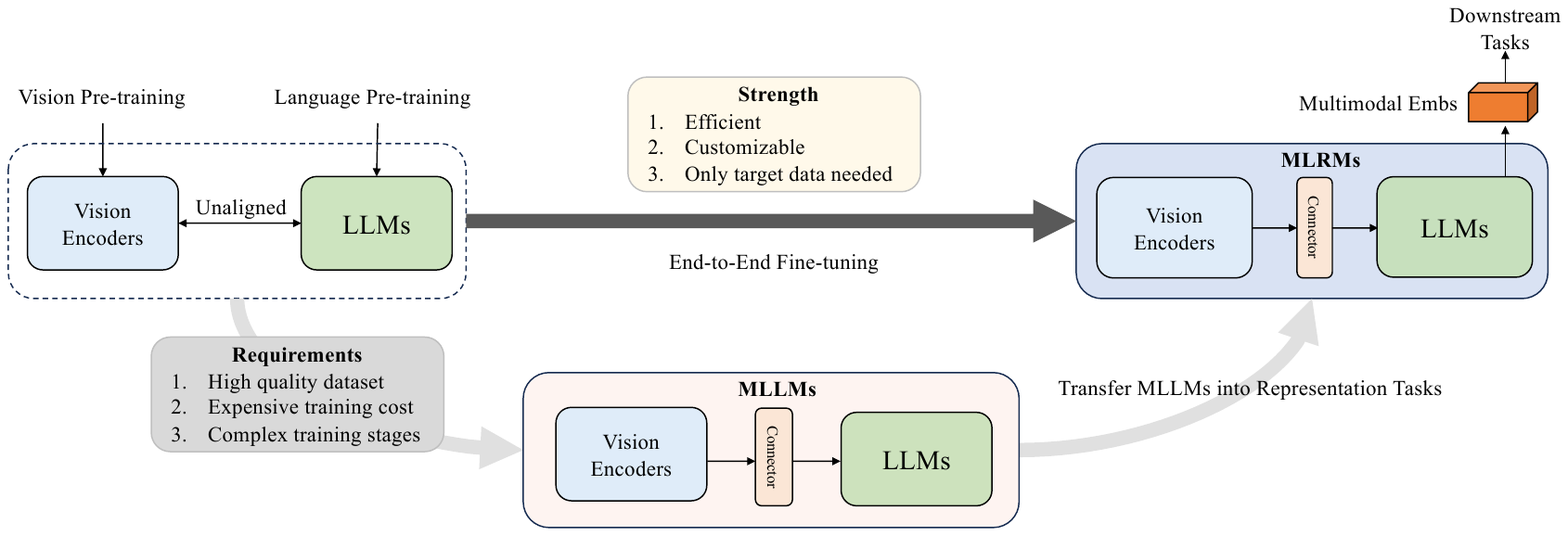}
    \caption{Two strategies for using LLMs to enhance multimodal representations.
    The first strategy transfers existing pre-trained MLLMs into representation tasks. The second strategy fine-tunes the integration of LLMs and vision encoders end-to-end, requiring no alignment and offering better practicality and efficiency. }
    \label{fig:route}
\end{figure*}
With the development of the internet, most online platforms are full of multimodal information to drive user engagement~\cite{wu2022mm,wei2019mmgcn,xun2021we}.
Multimodal recommendation is a necessary service for these platforms~\cite{liu2023multimodal}, which recommends items based on users' interest in item multimodal information.
These techniques mainly rely on multimodal representations, extracting raw multimodal content into dense embeddings~\cite{zhang2020multimodal}, and then measuring similarity through these multimodal embeddings.
The enhancement of multimodal representations can fundamentally improve numerous downstream tasks, such as retrieval~\cite{ma2023fine}, clustering~\cite{muennighoff2024generative}, and item modeling~\cite{wu2022mm}.

Many existing efforts have investigated multimodal representations, such as CLIP~\cite{radford2021learning} and METER~\cite{dou2022empirical}.
Despite significant success in these areas, the number of parameters in most multimodal representation models still has great potential for scaling~\cite{sun2023eva,sun2024eva}.
However, existing works~\cite{sun2023eva,sun2024eva} primarily focus on scaling vision models to improve multimodal representation, overlooking the textual branch.
At the same time, due to the lack of sufficient pure text-oriented pre-training~\cite{radford2021learning,sun2023eva,sun2024eva,zhai2023sigmoid}, the text branch in multimodal representation models often exhibits suboptimal understanding of text information~\cite{chen2023difference,zhang2024long}.
Therefore, it is necessary to enhance text understanding in multimodal representation models.


Recently, Large Language Models (LLMs) have shown impressive capabilities in text understanding and generating~\cite{jiang2023mistral,vicuna2023,bai2023qwen,touvron2023llama}.
Several works show the superior performance of LLMs in generating textual embeddings~\cite{muennighoff2024generative,zhang2024notellm,jiang2023scaling,ma2023fine,behnamghader2024llm2vec}.
However, the use of LLMs for enhancing text understanding in multimodal representations is underexplored.
As shown in Figure~\ref{fig:route}, one feasible method is to adapt existing open-source Multimodal Large Language Models (MLLMs)~\cite{liu2023improved,li2023blip,bai2023qwen,karamcheti2024prismatic} for representation tasks.
However, developing MLLMs requires substantial high-quality multimodal data and significant computational resources for pre-training~\cite{li2023blip,bai2023qwen}, resulting in the release of LLMs without multimodal capabilities ahead of their multimodal counterparts~\cite{deepseekai2024deepseekv2,jiang2023mistral,llama3}.
This delay constrains customized training using the most advanced LLMs and vision encoders.
Moreover, existing MLLMs are inefficient for representation scenarios as they primarily focus on multimodal understanding and generation~\cite{shang2024llava}.
Therefore, we design an end-to-end fine-tuning method that customizes the integration of any existing LLMs and vision encoders to construct efficient multimodal representation models.
This end-to-end method is practical when the target LLMs lack pre-trained visual components~\cite{deepseekai2024deepseekv2,jiang2023mistral,llama3}, as it only needs target domain data, eliminating pre-training data collection.
Besides, we can optimize the customized model structure for representation tasks to achieve better efficiency while ensuring performance.
We refer to these models for enhancing multimodal representation with LLMs as Multimodal Large Representation Models (MLRMs).
\begin{figure*}[!t]
    \centering
    \includegraphics[width=0.85\textwidth]{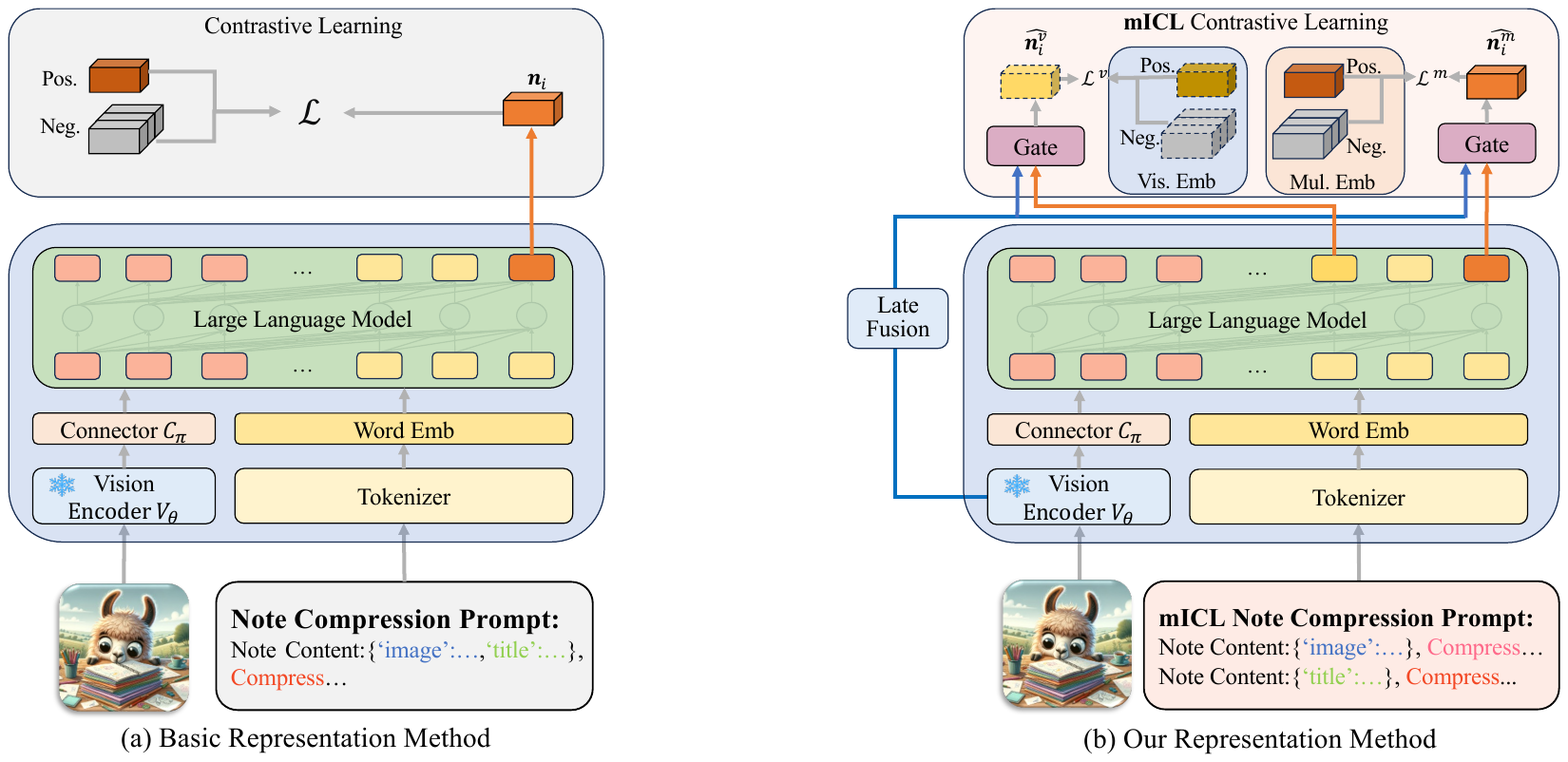}
    \vspace{-2.1mm}
    \caption{The frameworks of MLRMs. (a) is the basic method for representations. (b) is our representation method, which contains two easy and effective mechanisms to enhance multimodal representation ability, mICL and late fusion.}
    \vspace{-2.1mm}
    \label{fig:framework}
\end{figure*}

In this work, we explore MLRMs in item-to-item (I2I) multimodal recommendation scenarios.
We conduct preliminary fine-tuning experiments to examine our goal.
Our findings reveal that naive end-to-end fine-tuned MLRMs are biased towards text, ignoring visual content.
Images are mainly used to amplify key elements in multimodal content, rather than being directly aggregated by representations. 
To solve this challenge, we propose \method, a novel fine-tuning framework for multimodal tasks, with two methods to mitigate visual neglect.
The first method is multimodal In-Context Learning (mICL).
This method separates multimodal content into visual and textual components, subsequently compressing the content into two modality-compressed words.
Each modality-compressed word conducts in-batch contrastive learning with its corresponding modality-compressed word.
The second method is the late fusion mechanism.
Most MLLMs adopt early fusion, which maps visual features into the textual feature space of LLMs and interacts visual features with text embeddings through LLMs~\cite{liu2023improved,li2023blip,bai2023qwen,karamcheti2024prismatic}.
Even though early fusion can deepen the multimodal information interaction, it can easily lead to the loss of pure visual information during the multimodal fusion conducted by LLMs~\cite{li2023blip,zhang2024debiasing}.
We incorporate late fusion based on early fusion, retaining more visual information by delaying the fusion process.
Specifically, we adopt a multimodal gate fusion mechanism, which takes visual representations from the visual encoder and representations from the LLM to generate final multimodal representations.
We conduct extensive experiments to demonstrate our \method effectiveness.
Our paper makes the following contributions:
\begin{itemize}[leftmargin=*]
    \item To the best of our knowledge, we are the first to explore the multimodal representation assisted by LLMs in recommendation scenarios.
    We propose an end-to-end fine-tuning method that customizes the integration of any existing LLMs and vision encoders to form efficient MLRMs.
    \item 
    We reveal that fine-tuning unaligned MLRMs for representation tasks can cause modality imbalance by overlooking visual information.
    To overcome this problem, we design a novel framework, \method, which contains mICL and late fusion. 
    \item Extensive online and offline experiments have demonstrated the effectiveness of our proposed framework for multimodal recommendation representation.
\end{itemize}

\section{Preliminary Investigation}
\subsection{Problem Statement}
In our scenarios, every item represents a note, which is generated by users and expresses their life experiences.
We assume $\mathcal{N}=\{n_1,n_2,...,n_m\}$ as note pool, where $m$ is the number of notes.
Each note contains textual information, like the title, topic, and content, as well as the image. 
We denote the $i$-th note as $n_i=(t_i, tp_i, ct_i, v_i)$, where $t_i$, $tp_i$, $ct_i$, $v_i$ mean the title, the topic, the content and the image respectively.
Given a query note $n_i$, the I2I recommendation system ranks the note pool $\mathcal{N} \backslash \{n_i\}$ according to the multimodal content similarity to the given note.
The objective of this system is to prioritize the ranking of the related target note.

\subsection{Basic Representation Method of MLRMs}

Following~\cite{jiang2023scaling}, we adopt a prompt with an explicit one-word limitation to compress the multimodal content into one embedding.
Specifically, we use json format to show note content and our prompt is the following:
\begin{mdframed}
    \textbf{Note Compression Prompt} \\
    Note content: \{`image': <IMG>, `title': $t_i$ , `topic': $tp_i$ , `content': $ct_i$\}. Compress this note into one word:``
\end{mdframed}
In this template, <IMG> is a placeholder, which will be replaced by vision embeddings processed from raw images $v_i$.

The overall framework is shown in Figure~\ref{fig:framework} (a).
This framework first utilizes the vision encoder $V_{\theta}$ to extract the pre-processed image $v_i$ into visual features $Z_v\in \mathbb{R}^{L\times h_{v}}$ in the notes $n_i$, where $L$ is the length of visual features, $h_{v}$ is the dimension of visual features and $Z_v=V_{\theta}(v_i)$.
Then, the connector $C_{\pi}$ transforms the visual features $Z_v$ into the word embedding space of LLMs to form visual embeddings $E_v\in \mathbb{R}^{L_c\times h_{t}}$, where $L_c$ is the length of visual embeddings input into LLMs, $h_{t}$ is the dimension of LLMs' hidden states and $E_v = C_{\pi}(Z_v)$.
After processing images, the text prompts are tokenized into discrete indexes, then form the text word embedding $E_t\in \mathbb{R}^{T\times h_{t}}$, where $T$ is the length of text tokens.
To insert the visual embeddings into the correct position in text embeddings, we replace word embeddings in positions of the <IMG> token by the visual embeddings $E_v$, forming the multimodal embeddings $E_m\in \mathbb{R}^{(L_c+T-1)\times h_{t}}$.
Last, we use LLMs $LLM_{\mu}$ to process the multimodal embeddings to generate last hidden states $H\in \mathbb{R}^{(L_c+T-1)\times h_{t}}$, where $H=LLM_{\mu}(E_m)$.
We take the last embeddings of $H$ as the note representation $\boldsymbol{n}_i$ for note $n_i$.
This method constrains LLMs to compress the multimodal content into one embedding in the form of the next token prediction.

However, LLMs are trained using the language modeling loss~\cite{liu2023improved,li2023blip,bai2023qwenvl,karamcheti2024prismatic}, which significantly differs from representation tasks.
To bridge this gap, we employ contrastive learning, a prevalent technique for training embeddings~\cite{muennighoff2024generative,zhang2024notellm,jiang2023scaling,ma2023fine}.
Specifically, each minibatch contains $B$ related note pairs, resulting in a total of $2B$ notes per minibatch.
For any note $n_i$ (where $1 \leq i \leq 2B$) in the minibatch, we denote its related note in the same minibatch as $n_i^+$.
Following~\cite{neelakantan2022text}, we use gradient descent to minimize the contrastive loss as follows:
\begin{equation}
\mathcal{L}(\pi,\mu) =- \frac{1}{2B}\sum_{i=1}^{2B}log\frac{e^{sim(\boldsymbol{n}_i,\boldsymbol{n}_i^+)\cdot e^{\tau}}}{\sum_{j\in [2B]\backslash\{i\}}e^{sim(\boldsymbol{n}_i,\boldsymbol{n}_j)\cdot e^{\tau}}},
\end{equation}
where $\tau$ means the learnable temperature and $sim(a, b) = a^\top b /(\Vert a \Vert \Vert b \Vert)$.
$\mathcal{L}(\pi,\mu)$ means that we only update the connector $C_\pi$ and LLMs $LLM_\mu$, while keep the vision encoder $V_\theta$ frozen, aiming for the larger batch size and better performance~\cite{karamcheti2024prismatic}.

\begin{table}[!t]
    \centering
    
    \setlength\tabcolsep{2mm}
    \renewcommand\arraystretch{1.05}
    \caption{Detailed statistics of training and testing dataset.}
    \vspace{-2.1mm}
    \scalebox{0.8}{
    \begin{tabular}{l|r|l|r} \Xhline{1.0pt}
    \multicolumn{4}{c}{training dataset} \\
    \Xhline{1.0pt}
    \# notes & 1,487,613 & \# note pairs & 1,130,906 \\

    avg. \# words per title & 12.38 & avg. \# topic per note & 5.26\\

    avg. \# words per topic & 4.53 & avg. \# words per content & 133.20 \\
    
    \Xhline{1.0pt}
    \multicolumn{4}{c}{testing dataset} \\
    \Xhline{1.0pt}
    \# notes & 534,144 & \# note pairs & 21,262 \\

    avg. \# words per title & 11.68 & avg. \# topic per note & 4.58\\

    avg. \# words per topic & 4.50 & avg. \# words per content & 116.00 \\
    \Xhline{1.0pt}
    \end{tabular}}
    \label{tab:testingdatasets}
    \vspace{-3mm}
\end{table}

\begin{table*}[!h]
    \centering
    
    \setlength\tabcolsep{8pt}
    \renewcommand\arraystretch{0.9}
    \caption{Performance of different fine-tuned models (\%). ``\textbf{{\Large *}}'' indicates the statistically significant improvements (i.e., two-sided t-test with $p<0.05$) over the best baseline.}
    \vspace{-2.1mm}
    \scalebox{0.85}{
    \begin{tabular}{l|l|ccc|ccc|ccc} 
    \Xhline{1.0pt}
    \multirow{2}*{\textbf{Method}}&\multirow{2}*{\textbf{Input}}&\multicolumn{3}{c|}{\textbf{All Pair}}&\multicolumn{3}{c|}{\textbf{Short Query Pair}}&\multicolumn{3}{c}{\textbf{Short Target Pair}} \\
    \cline{3-11}
    ~&~&R@100&R@1k&R@10k&R@100&R@1k&R@10k & R@100&R@1k&R@10k \\
    \Xhline{0.5pt}
    \multicolumn{11}{c}{\textbf{Unimodal Training}} \\
    \Xhline{0.5pt}
    CLIP ViT-B&Image&23.21&45.44&71.16&25.24&49.40&75.99&25.58&50.28&75.80\\
    BM25&Text&55.22&72.64&82.82&35.75&48.41&59.81&35.05&49.52&63.64\\
    RoBERTa-wwm-ext&Text&67.59&85.06&93.86&41.23&58.61&76.40&41.80&60.65&80.04 \\
    Tomato&Text&71.01&87.38&95.19&43.55&60.70&79.10&45.04&63.45&83.24 \\
    Tomato + NoteLLM&Text&71.42&87.78&95.57&44.13&61.86&80.07&45.73&64.28&84.22 \\
    Qwen-Chat&Text&73.03&88.41&97.50&45.52&62.21&79.93&46.42&64.49&83.62 \\
    \Xhline{0.5pt}
    \multicolumn{11}{c}{\textbf{Multimodal Training}} \\
    \Xhline{0.5pt}
    METER Merge-attn&Multimodal&68.22&88.37&97.48&50.37&75.61&93.59&50.05&76.55&94.67 \\
    METER Co-attn&Multimodal&70.90&89.75&97.88&52.70&78.14&94.65&53.17&79.74&95.65 \\
    BLIP-2&Multimodal&68.38&88.22&97.44&52.97&77.83&94.40&54.00&79.74&95.29\\
    MTomato-Base&Multimodal&71.94&88.22&96.13&44.77&63.77&83.31&45.83&66.08&87.02\\
    MQwen-Base&Multimodal&74.02&89.65&97.22&48.15&68.35&88.22&49.82&70.91&91.43\\
    MQwen-bigG&Multimodal&77.63&92.89&98.91&57.45&80.92&95.73&57.95&83.50&96.99\\
    Qwen-VL-Chat\footnotemark{}&Multimodal&\textbf{78.53*}&\textbf{93.76*}&\textbf{99.03*}&\textbf{60.41*}&\textbf{83.27*}&\textbf{96.60*}&\textbf{61.60*}&\textbf{85.48*}&\textbf{97.89*}
\\
    \Xhline{1.0pt}
    \end{tabular}}
    \label{tab:fine-tune}
    \vspace{-2.1mm}
\end{table*}

\subsection{Datasets and Experimental Settings}

\textbf{Dataset Construction Method.}
To construct pairs of related notes without relying on human annotations, we leverage the co-occurrence mechanism~\cite{zhang2024notellm}, which measures the relevance between notes by how often they are read together.
We count the co-occurrence score in which users viewed note $n_A$ and subsequently clicked on note $n_B$ as follows:
\begin{equation}
s_{n_A\rightarrow n_B} = \sum_{u\in U_{n_A\rightarrow n_B}}\frac{1}{N_u},
\end{equation}
where $s_{n_A\rightarrow n_B}$ means the co-occurrence score from note $n_A$ to note $n_B$, $U_{n_A\rightarrow n_B}$ is the set of users who viewed note $n_A$ and subsequently clicked on note $n_B$, and $N_u$ denotes the number of distinct notes clicked by the user $u$ in the user behavior data.
We compute the co-occurrence score for each note pair.
Then, we construct the set $\mathcal{S}_{n_i}$, which contains co-occurrence scores from note $n_i$ to all other notes, i.e., $\mathcal{S}_{n_i}=\{{s_{n_i\rightarrow n_j}}|1\leq j \leq m, i\neq j\}$.
Next, we remove outlier from $\mathcal{S}_{n_i}$ with co-occurrence scores above $up$ or below $low$.
Finally, the top $t$ notes with the highest co-occurrence scores in the filtered set are viewed as the related notes for note $n_i$.

\label{para:exp_setting}
\textbf{Datasets Details.}
We obtain a real-world multimodal I2I dataset from our platform.
We create the training dataset by randomly choosing related note pairs from user behavior data gathered over two weeks.
One-tenth of the pairs in this training dataset are used for the validation set.
Next, the test note pool is collected by randomly selecting notes from the following week, while avoiding duplication with the training dataset.
Then, we count the related note pairs in the test note pool.
We provide detailed statistics of the training and testing dataset in Table~\ref{tab:testingdatasets}.

To better evaluate the multimodal representation ability of MLRMs rather than relying heavily on the text, we collect pairs that include short notes in the test dataset.
We define notes with a token length of less than $50$ as short notes, which approximately constitute $10\%$ of the total test notes.
We categorize pairs with short query notes as short query pairs and those with short target notes as short target pairs.
In the test dataset, the number of short query pairs is $5,620$ and the number of short target pairs is $5,582$.

\textbf{Experimental settings.}
When constructing pairs of related notes, the upper bound of the co-occurrence score $up$ is $30$ and the lower bound $low$ is $0.01$.
We set $t$ as $3$.
We truncate titles longer than 20 words and content longer than 80 words to comply with inference length restrictions.
In the fine-tuned experiments, for fair comparison, we add a linear projector to diminish the dimension of note embeddings to $64$. 
The batch size $B$ is set to $128$\footnote[1]{Models trained on 8 $\times$ 80GB Nvidia A100 GPUs and the batch size per GPU is $16$.}.
There are $256$ notes in each batch.
We start with a temperature $\tau$ set as $3$.
We provide more training details in Appendix~\ref{para:Training Details}.

To evaluate the representation ability for recall tasks, we create a similarity ranked list of all remaining notes in the test pool based on the query note's content.
We compute recall scores based on the position of the target note in the ranked list.
We report Recall@100, Recall@1k, and Recall@10k on all pairs, short query pairs, and short target pairs in the test dataset.
For robustness, we report the average results tested three times using random seeds \{42,43,44\}.

\subsection{Performance of Multimodal Representation of Fine-tuned MLRMs}

We conduct zero-shot experiments on several existing MLLMs in Appendix~\ref{para:zero-shot}, and find that zero-shot is insufficient to adapt MLLMs to representation tasks, performing inferiorly compared to a simple baseline, BM25~\cite{robertson2009probabilistic}.
This is due to the misalignment between pre-training tasks and representation tasks~\cite{zhang2024notellm,jiang2023scaling}.
Hence, fine-tuning MLLMs for representation tasks is necessary.

\begin{table*}[!h]
    \centering
    
    \setlength\tabcolsep{8pt}
    \renewcommand\arraystretch{0.9}
    \caption{Performance of MLRMs fine-tuned with multimodal inputs under different modal inputs (\%).}
    \vspace{-2.1mm}
    \scalebox{0.89}{
    \begin{tabular}{l|l|ccc|ccc|ccc} 
    \Xhline{1.0pt}
    \multirow{2}*{\textbf{Method}}&\multirow{2}*{\textbf{Input}}&\multicolumn{3}{c|}{\textbf{All Pair}}&\multicolumn{3}{c|}{\textbf{Short Query Pair}}&\multicolumn{3}{c}{\textbf{Short Target Pair}} \\
    \cline{3-11}
    ~&~&R@100&R@1k&R@10k&R@100&R@1k&R@10k & R@100&R@1k&R@10k \\
    \Xhline{0.5pt}
    \multicolumn{11}{c}{\textbf{Multimodal Training}} \\
    \Xhline{0.5pt}
    \multirow{3}*{BLIP-2}&Image&24.72&45.21&69.68&26.07&48.78&75.59&27.59&49.78&75.97 \\
    ~&Text&56.85&75.56&88.26&32.85&47.28&64.66&34.88&51.68&72.98 \\
    ~&Multimodal&68.38&88.22&97.44&52.97&77.83&94.40&54.00&79.74&95.29\\
    \Xhline{0.5pt}
    \multirow{3}*{MTomato-Base}&Image&0.60&1.49&6.98&0.54&1.51&9.15&0.84&2.13&9.70\\
    ~&Text&71.49&87.59&95.31&43.80&61.74&79.26&44.72&63.89&83.26 \\
    ~&Multimodal&71.94&88.22&96.13&44.77&63.77&83.31&45.83&66.08&87.02\\
    \Xhline{0.5pt}
    \multirow{3}*{MQwen-Base}&Image&1.92&6.59&23.43&2.86&8.54&28.83&3.05&9.38&29.72\\
    ~&Text&73.14&88.35&95.34&44.50&61.59&78.78&46.12&64.41&83.97 \\
    ~&Multimodal&74.02&89.65&97.22&48.15&68.35&88.22&49.82&70.91&91.43\\
    \Xhline{0.5pt}
    \multirow{3}*{MQwen-bigG}&Image&18.32&38.50&64.56&19.76&41.56&71.32&21.30&42.72&71.54\\
    ~&Text&70.63&86.09&93.81&39.94&55.77&72.17&43.03&60.02&79.77 \\
    ~&Multimodal&77.63&92.89&98.91&57.45&80.92&95.73&57.95&83.50&96.99\\
    \Xhline{0.5pt}
    \multirow{3}*{Qwen-VL-Chat}&Image&24.67&45.51&70.08&25.94&48.32&75.32&26.94&49.59&75.67
\\
    ~&Text&71.44&86.78&94.43&42.55&58.59&75.86&43.97&61.11&80.17
 \\
    ~&Multimodal&78.53&93.76&99.03&60.41&83.27&96.60&61.60&85.48&97.89
\\
    \Xhline{1.0pt}
    \end{tabular}}
    \label{tab:fine-tune diff modal}
    \vspace{-2.1mm}
\end{table*}

We design three end-to-end MLRMs to examine the representation fine-tuning method:
\textbf{MTomato-Base} uses Tomato (our continually pre-trained LLM based on LLaMA 2~\cite{touvron2023llama}, which lacks vision perception ability. See Appendix~\ref{para:MLRMs} for more details on pre-training.) as the LLM, CLIP ViT-B~\cite{radford2021learning} as the vision encoder and a randomly initialized Q-Former~\cite{li2023blip} as the connector for efficiency.
\textbf{MQwen-Base} replaces Tomato with Qwen-Chat~\cite{bai2023qwen} in MTomato-Base.
\textbf{MQwen-bigG} replaces CLIP ViT-B with ViT-bigG~\cite{ilharco_gabriel_2021_5143773} in MQwen-Base.
We set the visual embedding length of these models as $16$ for efficiency.
For comparison, we choose two pre-trained MLLMs: \textbf{BLIP-2} and \textbf{Qwen-VL-Chat}~\cite{bai2023qwenvl}.
All vision encoders in these models are frozen for the larger batch size.
We provide details about end-to-end MLRMs in Appendix~\ref{para:MLRMs}.

To evaluate the effectiveness of MLRMs, we compare them with several baselines.
(1) \textbf{CLIP ViT-B}~\cite{radford2021learning} is trained using only image input.
(2) \textbf{BM25}~\cite{robertson2009probabilistic} is a basic baseline, which counts the relevance between the query and documents based on every word in the query.
(3) \textbf{RoBERTa-wwm-ext}~\cite{cui2021pre} is a classic traditional Chinese BERT-like~\cite{kenton2019bert} text encoder with pure text input.
(4) \textbf{Tomato} is our continually pre-trained LLM that is based on LLaMA 2~\cite{touvron2023llama}.
(5) \textbf{NoteLLM}~\cite{zhang2024notellm} unifies the embedding generation and topic generation into one task.
(6) \textbf{METER Merge-attn} and \textbf{METER Co-attn}~\cite{dou2022empirical} leverage best training strategy to train vision-and-language transformers.
METER Merge-attn adopts the merged attention as the fusion module, while METER Co-attn uses co-attention as the fusion module.
All baselines are trained on our training dataset.

The results are shown in Table~\ref{tab:fine-tune}.
Several observations can be made.
Firstly, MLRMs can significantly outperform existing non-LLM-based baselines.
Qwen-VL-Chat achieves a $10.78\%$ improvement in R@100 compared to the traditional method, METER Co-attn on all pairs.
Secondly, the end-to-end fine-tuned representation method can enhance the model's multimodal representation ability.
MQwen-bigG improves R@100 on all pairs by $6.31\%$ compared to Qwen-Chat.
However, when the visual encoder is small, such as CLIP ViT-B, the enhancement in multimodal perception is not substantial.
MTomato-Base only improves R@100 on all pairs by $1.31\%$ compared to Tomato, and MQwen-Base only improves R@100 on all pairs by $1.36\%$ compared to Qwen-Chat.
Lastly, although MQwen-bigG is more efficient than Qwen-VL-Chat, which uses the same vision encoder and LLM (see Section~\ref{para:comparison}), there remains a performance gap.

We also evaluate all MLRMs fine-tuned with multimodal inputs under different modal inputs, as shown in Table~\ref{tab:fine-tune diff modal}.
Firstly, we observe that MTomato-Base and MQwen-Base hardly represent the image content when the input consists solely of images, which ignores visual information.
Owing to its more robust vision encoder, MQwen-bigG exhibits superior vision representation capabilities.
However, its performance still falls short when compared to BLIP-2 and Qwen-VL-Chat, both of which have undergone multimodal pre-training.
\setcounter{footnote}{1}
\footnotetext{Due to Qwen-VL-Chat having $256$ visual embeddings per image, this model is trained on 32 $\times$ 80GB Nvidia A100 GPUs and the batch size per GPU is $4$.}

\subsection{Exploring the Information Flow of Fine-tuned MLRMs}

\begin{figure*}[!t]
    \centering
    \subfigure[BLIP-2]{
    \includegraphics[width=0.23\textwidth]{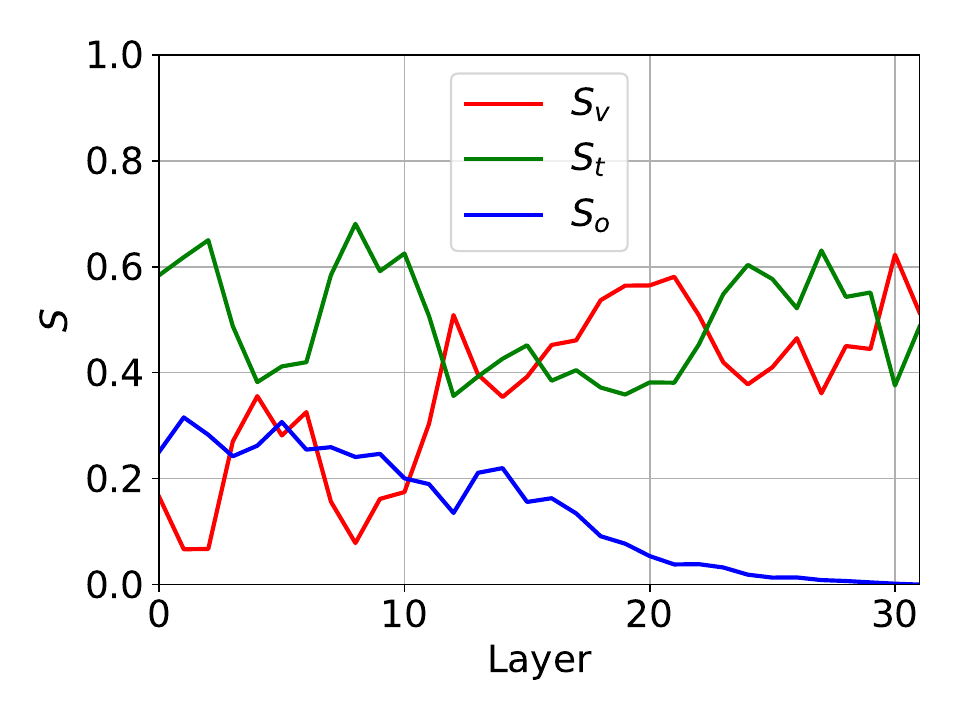}}
    \subfigure[MTomato-Base]{
    \includegraphics[width=0.23\textwidth]{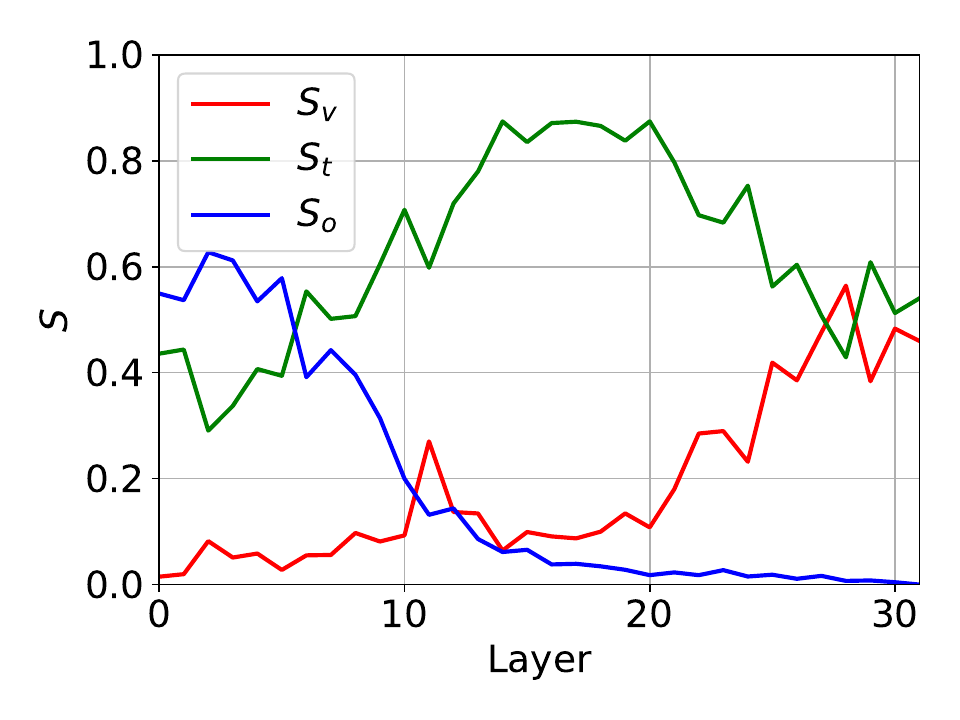}}
    \subfigure[MQwen-Base]{
    \includegraphics[width=0.23\textwidth]{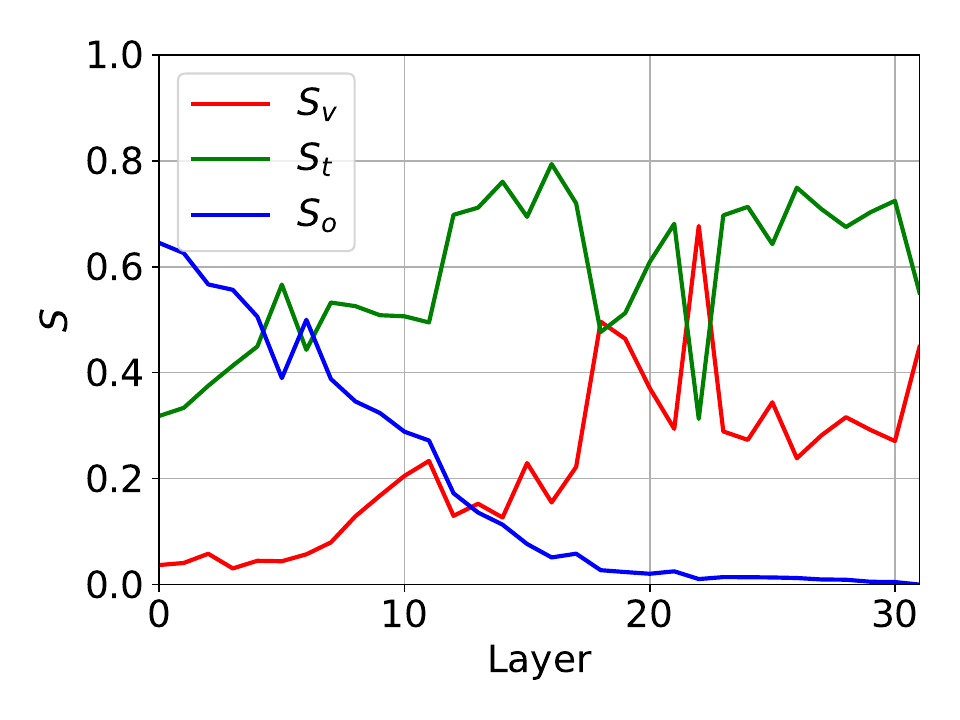}}
    \subfigure[MQwen-bigG]{
    \includegraphics[width=0.23\textwidth]{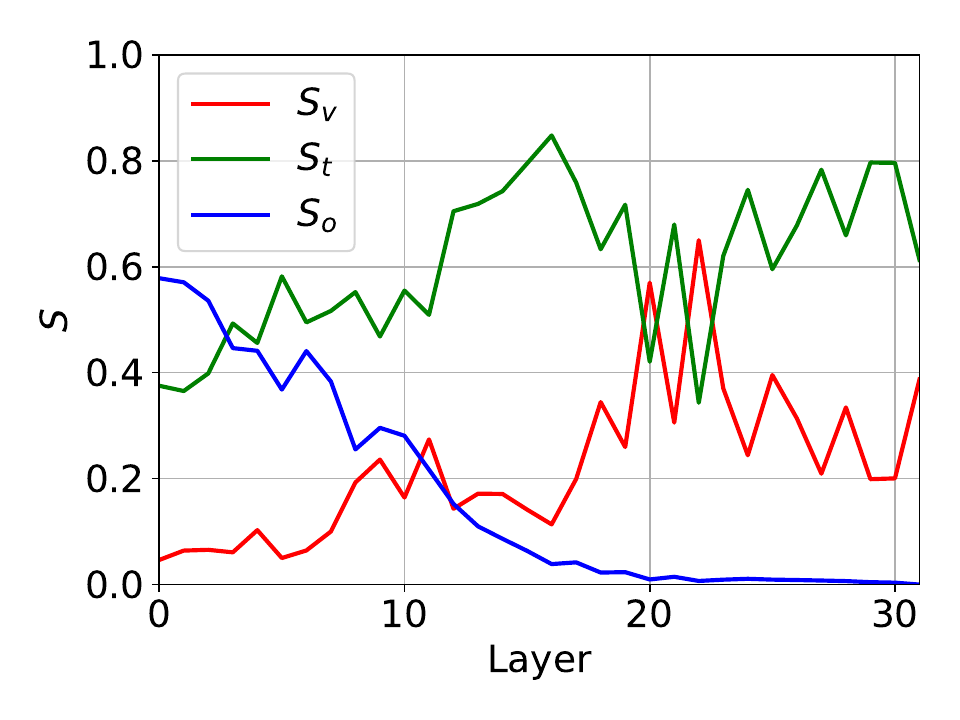}}
    \vspace{-3mm}
    \caption{Relative sizes of $S_v$, $S_t$, and $S_o$ in different layers of different MLRMs.
    $S_v$ and $S_t$ are the mean significance of information flows from visual embeddings and textual embeddings to final compressed embeddings, respectively.
    $S_o$ is the mean significance of information flow among all embeddings except final compressed embeddings.}
    \label{fig:saliency_score}
    \vspace{-2.1mm}
\end{figure*} 

To further explore the reasons why end-to-end fine-tuned MLRMs overlook visual information, we examine the attention patterns of two modality tokens in different LLMs' layers from the perspective of information flow.
We adopt saliency scores~\cite{wang2023label} to analyze the interaction between the compressed word and two modality tokens in the attention matrices.
We randomly select $10,000$ notes in batches from the training dataset to count the saliency scores,
and compute the saliency scores as follows:
\begin{equation}
I_l = \sum_h \left| A^{\top}_{h,l} \frac{\partial \mathcal{L}}{\partial A_{h,l}} \right|,
\end{equation}
where $A_{h,l}$ means the attention matrix value from the $h$-th attention head within $l$-th layer and $\mathcal{L}$ is the contrastive loss.
The saliency matrix $I_l$ for layer $l$ is computed by summing across all attention heads.
$I_l(i,j)$ means the significance of the information flow from the $j$-th word to the $i$-th word.
To analyze the relative importance of different modalities for the compressed word, we separate $I_l$ into three parts as follows:
$S_v$, the mean significance of information flow from the vision embeddings $E_v$ to the final compressed position $c$:
\begin{equation}
\begin{aligned}
S_v &= \frac{\sum_{(i,j)\in P_v}I_l(i,j)}{\left| P_v \right|}, \\
P_v &= \{(c,j): j\in E_v \}.
\end{aligned}
\end{equation}

$S_t$, the mean significance of information flow from the textual embeddings $E_t$ to the final compressed position $c$:
\begin{equation}
\begin{aligned}
S_t &= \frac{\sum_{(i,j)\in P_t}I_l(i,j)}{\left| P_t \right|}, \\
P_t &= \{(c,j): j\in E_t , j<c \}.
\end{aligned}
\end{equation}

$S_o$, the mean significance of information flow amongst all words, excluding influences represented by $S_v$ and $S_t$:
\begin{equation}
\begin{aligned}
S_o &= \frac{\sum_{(i,j)\in P_o}I_l(i,j)}{\left| P_o \right|}, \\
P_o &= \{(i,j): j<i \}-P_v-P_t.
\end{aligned}
\end{equation}

The results of three parts from the saliency matrix are shown in Figure~\ref{fig:saliency_score}.
We observe that pre-trained MLLMs, BLIP-2 and Qwen-VL-Chat (see Appendix~\ref{para:qwsaliency score}), have balanced $S_v$ and $S_o$ in shallow layers.
However, MLRMs fine-tuned end-to-end without pre-training exhibit a low $S_v$ and a high $S_o$ in shallow layers.
The low $S_v$ indicates limited direct information flow from visual to final compressed embeddings, while the high $S_o$ indicates significant information flow between visual and textual embeddings.
This shows that the information flow gain brought by introducing images is more likely to flow towards important multimodal embeddings, rather than being directly aggregated into final compressed embeddings.
As the number of layers increases, we observe that $S_v$ increases while $S_o$ decreases dramatically.
This is because, after multimodal embeddings capture context information in shallow layers (as shown by $S_o$), models focus on directly aggregating relevant information into final compressed embeddings for representation tasks.

Unlike pre-trained MLLMs, the final compressed embeddings of end-to-end MLRMs have little direct visual information in shallow layers.
This causes the final compressed embedding initially accumulates visual information mainly through the aggregation of context information contained in textual embeddings.
In deep layers, final compressed embeddings increasingly incorporate image information related to the previously aggregated textual information through the attention mechanism.
Therefore, images in end-to-end MLRMs primarily assist the information flow in identifying crucial information~\cite{sun2021rpbert,sun2020riva}, rather than being directly aggregated by representations.
Besides, text information flow significantly surpasses visual information flow in end-to-end fine-tuned MLRMs.

\begin{table*}[!h]
    \centering
    
    \setlength\tabcolsep{8pt}
    \renewcommand\arraystretch{0.9}
    \caption{Performance of different methods based on three MLRMs (\%). ``\textbf{{\Large *}}'' indicates the statistically significant improvements (i.e., two-sided t-test with $p<0.05$) over the best baseline.}
    \scalebox{0.85}{
    \begin{tabular}{l|l|ccc|ccc|ccc} 
    \Xhline{1.0pt}
    \multirow{2}*{\textbf{Base Model}}&\multirow{2}*{\textbf{Method}}&\multicolumn{3}{c|}{\textbf{All Pair}}&\multicolumn{3}{c|}{\textbf{Short Query Pair}}&\multicolumn{3}{c}{\textbf{Short Target Pair}} \\
    \cline{3-11}
    ~&~&R@100&R@1k&R@10k&R@100&R@1k&R@10k & R@100&R@1k&R@10k \\
    \Xhline{0.5pt}
    \multirow{6}*{MTomato-Base}&basic method&71.94&88.22&96.13&44.77&63.77&83.31&45.83&66.08&87.02\\
    ~&mICL&74.15&90.94&98.26&51.56&75.87&93.84&52.74&78.13&\textbf{95.82} \\
    ~&late fusion&74.51&91.37&98.45&54.19&78.02&94.52&54.00&79.71&95.80 \\
    ~&\method&\textbf{74.62*}&\textbf{91.50*}&\textbf{98.47}&\textbf{54.87*}&\textbf{78.82*}&\textbf{94.69*}&\textbf{54.51*}&79.75&95.46 \\
    ~&only late fusion&74.23&91.25&98.43&53.20&77.75&94.65&53.65&\textbf{79.76}&95.59 \\
    ~&Omni-Retrieval&71.73&88.12&96.32&44.98&64.26&84.57&45.18&66.14&88.48 \\
\Xhline{0.5pt}
\multirow{6}*{MQwen-Base}&basic method&74.02&89.65&97.22&48.15&68.35&88.22&49.82&70.91&91.43
\\
    ~& mICL&75.81&91.57&98.55&53.32&76.40&94.38&54.55&79.32&96.05
 \\
    ~&late fusion&75.92&91.50&98.28&52.74&76.03&93.65&54.06&78.89&95.53
 \\
 ~&\method &\textbf{76.49*}&\textbf{92.18*}&\textbf{98.57}&\textbf{55.73*}&\textbf{78.91*}&\textbf{94.94*}&\textbf{56.12*}&\textbf{80.65*}&\textbf{96.24*} \\
    ~&only late fusion&75.84&91.48&98.35&53.33&76.22&93.57&54.27&78.38&95.29 \\
    ~&Omni-Retrieval&74.71&91.05&98.11&51.61&74.34&92.90&51.91&77.07&94.71 \\
\Xhline{0.5pt}
    \multirow{6}*{MQwen-bigG}&basic method&77.63&92.89&98.91&57.45&80.92&95.73&57.95&83.50&96.99
\\
    ~&mICL&\textbf{77.96*}&93.18&\textbf{98.92}&\textbf{58.48*}&\textbf{82.09*}&\textbf{96.65*}&\textbf{58.76}&\textbf{84.01}&\textbf{97.20*}
 \\
    ~&late fusion&77.43&92.93&98.78&57.47&81.78&95.89&58.48&83.99&97.16
 \\
 ~& \method&77.56&\textbf{93.25}&98.82&57.94&81.98&96.29&\textbf{58.76}&83.87&97.05 \\
    ~&only late fusion&76.23&92.30&98.67&55.21&79.22&95.58&55.16&81.13&96.70
 \\
 ~&Omni-Retrieval&77.38&93.11&98.91&58.26&82.07&96.61&58.28&83.95&97.13 \\
    \Xhline{1.0pt}
    \end{tabular}}
    \label{tab:ablation}
\end{table*}

\section{Methodology}
From our investigation, MLRMs trained end-to-end without pre-training are suboptimal, primarily due to the disregard of visual information after processing by LLMs.
Therefore, we aim to design mechanisms that focus more on visual signals.
We propose a novel training framework, \textbf{\method}, which contains two methods from different perspectives.
The first one is from the prompt viewpoint and is called mICL.
This method alters the prompt to adjust the attention patterns towards visual information.
The second one is from the model architecture perspective, which combines the late fusion with the visual prompt.
This enhances the impact of the visual information flow on the final representation by delaying the fusion of visual information.
The improved overall framework is shown in Figure~\ref{fig:framework} (b).

Specifically, given note $n_i$, the mICL does not compress the multimodal information into one compressed word.
Instead, we separate the multimodal note into two single-modality notes.
We then adopt a similar ICL way~\cite{dong2022survey,wang2023label} to aggregate the multimodal information.
We reformulate our note compression prompt as follows:

\begin{mdframed}
    \textbf{mICL Note Compression Prompt.} \\
    Note content: \{`image': <IMG>\}, Compress this note into one word:``<IMG\_EMB>''. Note content: \{`title': $t_i$ , `topic': $tp_i$ , `content': $ct_i$\}. Compress this note into one word:`` 
\end{mdframed}
In this template, <IMG\_EMB> is a special token.

After processing the multimodal embeddings with LLMs, we choose relevant hidden states to represent notes.
Following~\cite{zhang2024notellm,jiang2023scaling}, we take the previous hidden state of the token <IMG\_EMB> as the visual note representation, denoted as $\boldsymbol{n}_i^v$.
Due to the causal attention, $\boldsymbol{n}_i^v$ only contains the note image information.
We also take the last hidden state as the multimodal note representation, denoted as $\boldsymbol{n}_i^m$, which contains visual and textual information.
This method separates the original single embedding into two modality note embeddings.
The multimodal note representation $\boldsymbol{n}_i^m$ aggregates useful visual information in a similar manner to ICL.

The late fusion adopts original visual embedding to enhance the note embeddings.
This prevents a textual bias due to the space of LLMs~\cite{zhang2024debiasing} and incorporates more original visual information.
Specifically, after extracting image features by the vision encoder $V_\theta$, we take the visual feature, which contains the entire image information, such as the [CLS] token in CLIP ViT-B.
Then, we use a linear layer to transform these features into the dimension of LLMs, denoted as $\boldsymbol{v}\in \mathbb{R}^{h_t}$.
We adopt the gate mechanism~\cite{chung2014empirical} to fuse the original visual information into the note embeddings.
\begin{equation}
\begin{aligned}
\boldsymbol{z} &= sigmoid(\boldsymbol{W}[\boldsymbol{v},\boldsymbol{n}_i^v]+\boldsymbol{b}), \\
\hat{\boldsymbol{n}_i^v} &= \boldsymbol{z} \odot \boldsymbol{v} + (\boldsymbol{1}-\boldsymbol{z}) \odot \boldsymbol{n}_i^v,
\end{aligned}
\end{equation}
where $\hat{\boldsymbol{n}_i^v}$ means the fused visual note embeddings.
$[\cdot,\cdot]$ means the concatenate operation, and $\boldsymbol{W}$ and $\boldsymbol{b}$ are learnable parameters.
$\odot$ is the element-wise product.
Similarly, we can get the fused multimodal note embedding $\hat{\boldsymbol{n}_i^m}$ with the same gate operation.

Next, we adopt the fused note embeddings to conduct contrastive learning as follows:
\begin{equation}
\mathcal{L}^v(\pi,\mu) =- \frac{1}{2B}\sum_{i=1}^{2B}log\frac{e^{sim(\hat{\boldsymbol{n}_i^v},\hat{\boldsymbol{n}_i^{v,+}})\cdot e^{\tau}}}{\sum_{j\in [2B]\backslash\{i\}}e^{sim(\hat{\boldsymbol{n}_i^v},\hat{\boldsymbol{n}_j^v})\cdot e^{\tau}}},
\end{equation}
where $\mathcal{L}^v(\pi,\mu)$ is the loss from visual note embeddings. Similarly, we can get $\mathcal{L}^m(\pi,\mu)$, which is the loss from the multimodal note embeddings.
The final loss is computed as follows:
\begin{equation}
\mathcal{L}^f(\pi,\mu) = \frac{\mathcal{L}^v(\pi,\mu) + \alpha \mathcal{L}^m(\pi,\mu)}{1+\alpha},
\end{equation}
where $\mathcal{L}^f(\pi,\mu)$ is the final loss and $\alpha$ is a hyperparameter.
In the evaluation, we use $\hat{\boldsymbol{n}_i^m}$ as the note embeddings, which contain multimodal information.

\section{Experiments}
\begin{figure*}[!t]
    \centering
    \subfigure[MTomato-Base]{
    \includegraphics[width=0.28\textwidth]{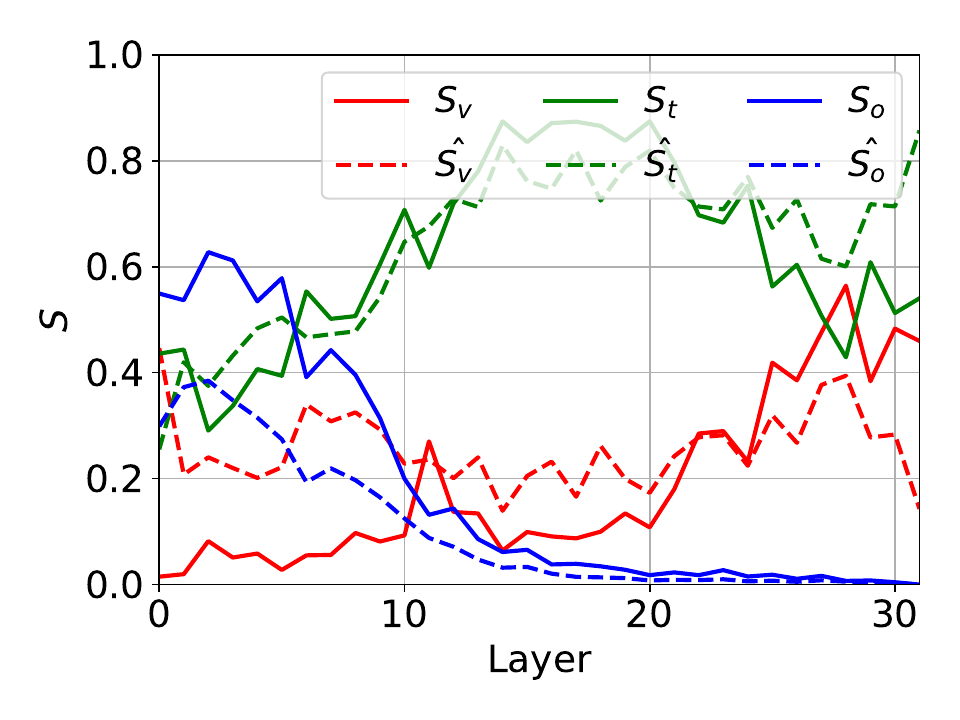}}
    \subfigure[MQwen-Base]{
    \includegraphics[width=0.28\textwidth]{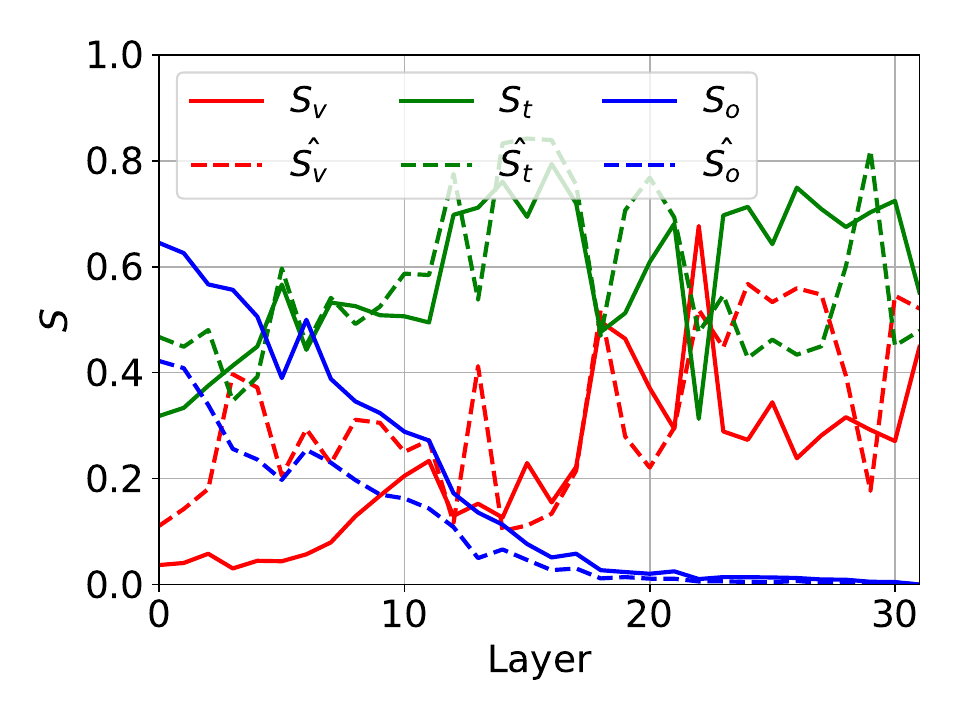}}
    \subfigure[MQwen-bigG]{
    \includegraphics[width=0.28\textwidth]{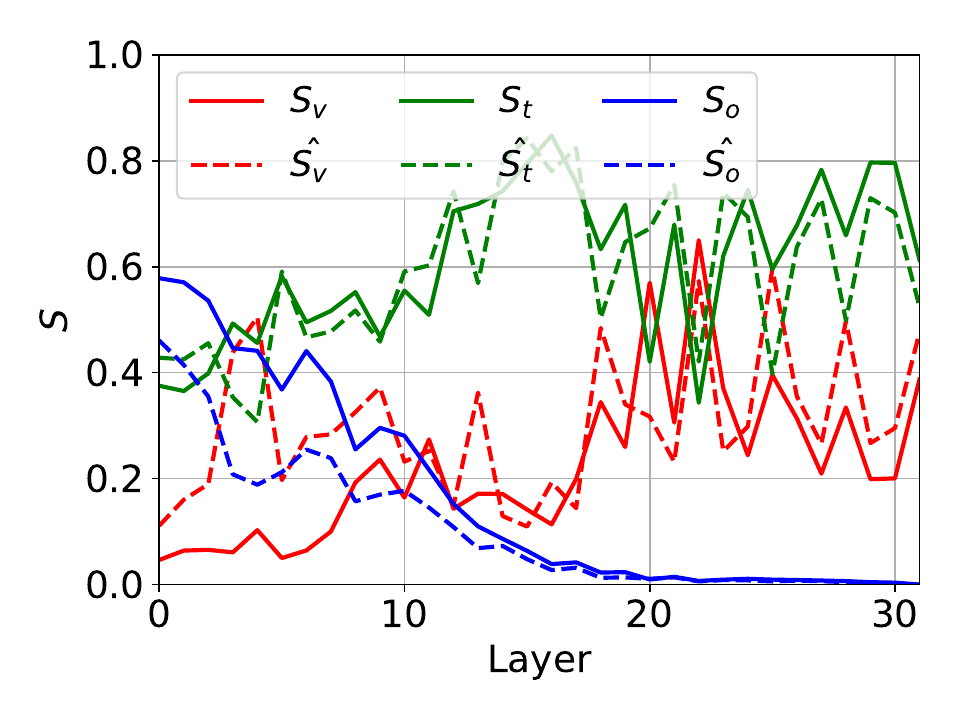}}
    \vspace{-2.1mm}
    \caption{Relative sizes of saliency scores in different layers of MLRMs. $S_v$ and $S_t$ are the mean significance of information flows from visual and textual embeddings to final compressed embeddings, respectively.
    $S_o$ is the mean significance of information flow among all embeddings except final compressed embeddings. 
    $\hat{S_v}$, $\hat{S_t}$, and $\hat{S_o}$ are saliency scores enhanced by \method.}
    \label{fig:ssdif}
\end{figure*}

\subsection{Performance Evaluation}
We conduct experiments on following baselines based on three MLRMs to show \method's effectiveness.
\textbf{basic method} is the naive representation method of MLRMs.
\textbf{mICL} adds the mICL method based on the basic method.
\textbf{late fusion} integrates the late fusion mechanism into the basic method.
\textbf{\method} contains both mICL and late fusion simultaneously.
\textbf{only late fusion} only uses late fusion to integrate image and text information, without inputting image embeddings into LLMs.
\textbf{Omni-Retrieval}~\cite{yu2022commercemm} adopts cross-modal contrastive learning across various embeddings, including image-only, text-only, and multimodal, to achieve alignment.

The results are presented in Table~\ref{tab:ablation}.
Firstly, \method can significantly enhance the overall performance of MTomato-Base and MQwen-Base, which have relatively small vision encoders.
Additionally, \method primarily enhances MQwen-bigG in terms of short pairs.
Secondly, mICL enhances the performance of all models, while late fusion is more effective in models with relatively small vision encoders.
When the vision encoders are powerful, MQwen-bigG+late fusion improves performance on short pairs by focusing more on images.
However, this overemphasis on images can detrimentally affect text comprehension, reducing overall performance and weakening the mICL method that utilizes LLM prompts.
We will design a more balanced late fusion mechanism in future work.
Concurrently, only late fusion is a straightforward and efficient fusion method.
However, its simplicity could potentially be a drawback when the vision encoder is more powerful, as it may not fully interact with LLMs, leading to a loss in performance.
Besides, the enhancement of Omni-Retrieval for visual information is still limited compared to \method. This is because Omni-Retrieval ignores the distinct mechanisms of LLMs, and other cross-modal losses may interfere with multimodal embeddings.
Lastly, MQwen-bigG+\method underperforms compared to Qwen-VL-Chat, especially in shorter pairs.
This is due to the visual token length difference between MQwen-bigG ($16$) and Qwen-VL-Chat ($256$).

\subsection{Saliency Scores of Enhanced MLRMs}
To further understand the impact of \method on MLRMs, we illustrate the difference in saliency scores of enhanced MLRMs.
The saliency scores for the original fine-tuned method are $S_v$, $S_t$, and $S_o$.
The saliency scores for the enhanced fine-tuned method are $\hat{S_v}$, $\hat{S_t}$ and $\hat{S_o}$.
We consider the visual note compressed word as a part of the vision embeddings $E_v$.
Our results are shown in Figure~\ref{fig:ssdif}.
The representations of all enhanced MLRMs amplify $S_v$, the direct focus on images, while reducing the emphasis on $S_o$ in the shallow layers.
At the same time, $S_t$ remains virtually unchanged.
More balanced $S_v$ and $S_o$ mean that the information flow gain brought by introducing images not only highlights the relevant multimodal embeddings but is also aggregated by the final compressed embedding directly. 
This is helpful for final compressed embeddings to capture crucial patterns in images that are totally different from text.
This result is attributed to mICL's use of the same compressed prompt to compress both modalities.
By identifying similar compression patterns from images, mICL can bolster the concentration of multimodal representations on images, similar to ICL~\cite{wang2023label}.

\subsection{Hyper-Parameter Analysis}
In this section, we use MTomato-Base+\method to conduct hyper-parameter analysis experiments.

\textbf{Impact of length of visual tokens.}
We explore the impact of visual token length.
The results are shown in Table~\ref{tab:vlength}.
Reducing the length from $16$ to $8$ leads to poorer performance, while inference is more efficient.
There exists a tradeoff between performance and inference speed concerning the length of visual tokens.
Moreover, MTomato-Base+\method with 8 visual tokens outperforms MTomato-Base with 16 visual tokens in Table~\ref{tab:fine-tune}.
This shows that our method enhances visual attention by modified attention patterns and fusion, rather than merely expanding image token space.

\textbf{Impact of text and visual loss ratio.}
We vary the ratio of text to visual loss $\alpha$, as shown in Table~\ref{tab:ratio}.
When the ratio is small, the performance is slightly worse.
However, as we increase the ratio, we find that our method is insensitive to changes in the text to visual loss ratio.

\begin{table}[!t]
    \centering
    
    \setlength\tabcolsep{2pt}
    \renewcommand\arraystretch{0.9}
    \caption{Impact of the number of visual tokens $L_c$ (\%). Inf. Sp. means the inference speed.}
    \scalebox{0.75}{
    \begin{tabular}{m{0.6cm}<{\centering}|c|ccc|ccc|ccc} 
    \Xhline{1.0pt}
    \multirow{2}*{\textbf{$L_c$}}&\multirow{2}*{\textbf{Inf. Sp.}}&\multicolumn{3}{c|}{\textbf{All Pair}}&\multicolumn{3}{c|}{\textbf{Short Query Pair}}&\multicolumn{3}{c}{\textbf{Short Target Pair}} \\
    \cline{3-11}
    ~&~&R@100&R@1k&R@10k&R@100&R@1k&R@10k & R@100&R@1k&R@10k \\
    \Xhline{0.5pt}
    8&51.0&74.17&90.85&98.16&52.80&76.13&93.76&53.19&77.87&94.48 \\
    16&48.2&\textbf{74.62}&91.50&98.47&\textbf{54.87}&78.82&94.69&\textbf{54.51}&79.75&95.46 \\
    32&44.2&74.55&91.41&98.38&54.56&\textbf{78.91}&94.94&54.21&79.83&95.63 \\
    48&41.0&74.30&\textbf{91.57}&\textbf{98.48}&54.31&78.83&\textbf{95.13}&53.73&\textbf{80.06}&\textbf{95.92} \\
    \Xhline{1.0pt}
    \end{tabular}}
    \label{tab:vlength}
\end{table}

\begin{table}[!t]
    \centering
    
    \setlength\tabcolsep{2pt}
    \renewcommand\arraystretch{0.9}
    \caption{Impact of textual loss and visual loss ratio $\alpha$ (\%).}
    \scalebox{0.8}{
    \begin{tabular}{m{1cm}<{\centering}|ccc|ccc|ccc} 
    \Xhline{1.0pt}
    \multirow{2}*{\textbf{$\alpha$}}&\multicolumn{3}{c|}{\textbf{All Pair}}&\multicolumn{3}{c|}{\textbf{Short Query Pair}}&\multicolumn{3}{c}{\textbf{Short Target Pair}} \\
    \cline{2-10}
    ~&R@100&R@1k&R@10k&R@100&R@1k&R@10k & R@100&R@1k&R@10k \\
    \Xhline{0.5pt}
    1 &73.94&91.11&98.38&54.13&78.16&94.67&54.06&79.00&95.38 \\
    3&74.38&91.41&98.42&54.15&77.71&94.59&54.06&79.31&\textbf{95.53} \\
    9&\textbf{74.62}&\textbf{91.50}&98.47&\textbf{54.87}&\textbf{78.82}&94.69&\textbf{54.51}&79.75&95.46 \\
    19&74.53&91.33&\textbf{98.54}&54.56&78.68&\textbf{95.00}&53.92&\textbf{80.04}&\textbf{95.53} \\
    \Xhline{1.0pt}
    \end{tabular}}
    \label{tab:ratio}
\end{table}

\begin{figure}[t]
    \centering
    \includegraphics[width=0.49\textwidth]{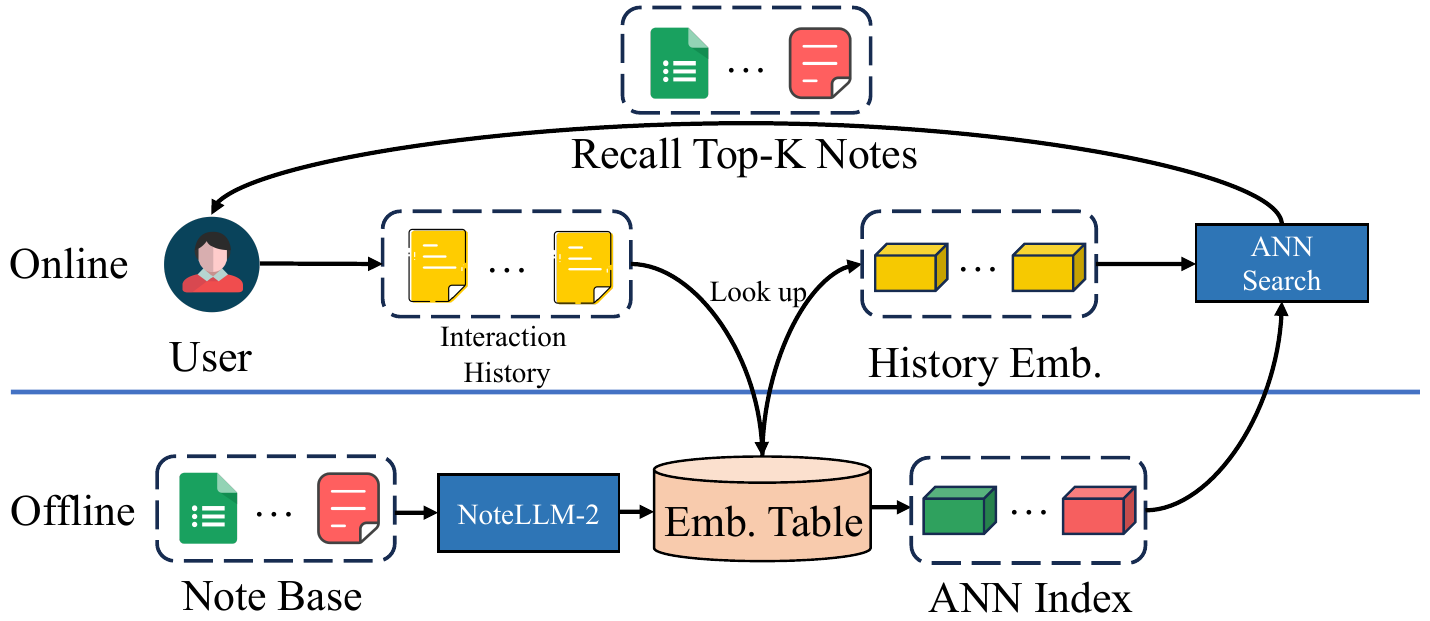}
    \caption{The online serving pipeline includes offline and online phases. The offline phase generates embeddings with the MLRM trained with \method and updates the ANN index with new notes. The online phase recalls notes from the index based on the user's interaction history. }
    \label{fig:online}
\end{figure}

\begin{table*}[!t]
    \centering
    
    \setlength\tabcolsep{7pt}
    \renewcommand\arraystretch{0.9}
    \caption{Comparison of different LLMs and MLRMs for representation tasks. $L_c$ means the number of vision embeddings for each image. All models are trained on A100 GPUs. The training cost is measured in GPU hours (GPU-h). The inference speed test is conducted on an A100 GPU, with an inference batch size set at $50$. All notes in this test are the longest note in the dataset. The speed is quantified as the number of longest notes inferred per second.}
    \scalebox{0.85}{
    \begin{tabular}{l|l|lll|cccc} 
    \Xhline{1.0pt}
    \textbf{Method}&\textbf{Input}&\textbf{Vision Encoder}&\textbf{LLM}&\textbf{VL Connector}&\textbf{\# Total Par.}&\textbf{$L_c$}&\textbf{ Training GPU-h}&\textbf{Inf. Speed} \\
    \Xhline{0.5pt}
    Tomato & Text & - & Tomato (6.7B) & - & 6.7B & - & 78.3 & 59.0 \\
    Qwen-Chat & Text & - & Qwen-Chat (7.1B) & - & 7.1B & - & 86.0  & 57.0 \\
    \Xhline{0.5pt}
    BLIP-2&Multimodal&ViT-g (1.0B)&OPT (6.7B)&Q-Former (105.1M)&7.8B&32& 110.8 &44.2\\
    Qwen-VL-Chat&Multimodal&ViT-bigG (1.8B)&Qwen-Chat (7.1B)&Resamplers (76.1M)&9.0B&256&432.6&11.5\\
    \Xhline{0.5pt}
    MTomato-Base&Multimodal &ViT-B (85.8M)&Tomato (6.7B)&Q-Former (49.6M)&6.8B&16&108.6&51.5\\
    \quad + \method&Multimodal&ViT-B (85.8M)&Tomato (6.7B)&Q-Former (49.6M)&6.9B&16&110.7&48.2\\
    \Xhline{0.5pt}
    MQwen-Base&Multimodal &ViT-B (85.8M)&Qwen-Chat (7.1B)&Q-Former (49.6M)&7.3B&16&100.3&50.5\\
    \quad + \method&Multimodal&ViT-B (85.8M)&Qwen-Chat (7.1B)&Q-Former (49.6M)&7.3B&16&101.6&46.9\\
    \Xhline{0.5pt}
    MQwen-bigG&Multimodal &ViT-bigG (1.8B)&Qwen-Chat (7.1B)&Q-Former (142.9M)&9.1B&16&129.8&36.8\\
    \quad + \method&Multimodal&ViT-bigG (1.8B)&Qwen-Chat (7.1B)&Q-Former (142.9M)&9.2B&16&123.1&35.2\\
    \Xhline{1.0pt}
    \end{tabular}}
    \label{tab:sum}
\end{table*}

\subsection{Efficiency Analysis} 
We provide a comparison of the model parameter size, training cost, and inference speed of LLMs in Table~\ref{tab:sum}.
State-of-the-art MLLMs such as Qwen-VL-Chat, designed for multimodal understanding and answering, have high training costs for adaptation and slow inference speed, making them impractical for fast-paced recommendation tasks.
On the other hand, MLRMs designed for representation tasks are more efficient for both training and inference.
Additionally, due to \method's longer prompt template, integrating \method usually slightly slows down training and inference, but achieves better performance.

\label{para:comparison}

\subsection{Online Serving and Experiments}

We present our online serving pipeline in Figure~\ref{fig:online}, which consists of offline and online phases.
During the offline phase, the MLRM trained with \method generates multimodal embeddings in real-time for newly generated notes from users.
These embeddings are saved in an embedding table for reuse.
Furthermore, the I2I ANN index is dynamically updated with embeddings of new notes.
In the online phase, the system selects history embeddings from the existing embedding table based on the user's interaction history notes.
Using these history note embeddings, the system performs multiple ANN searches to find relevant notes in the I2I index.
The top-k relevant notes are recalled and passed to the next recommendation stage.
Now, our model has replaced the previous I2I recall process and has already become a crucial recall channel on the platform.

We conduct week-long online experiments on our platform by allocating 10\% of the overall traffic for A/B testing.
The online baseline uses the same LLM as our online method for I2I training, with the training approach utilizing NoteLLM~\cite{zhang2024notellm}.
Our MLRM trained with \method enhances the click count of the first thousand exposures by $6.35\%$ and the number of interactions within the first $24$ hours of note publication by $8.08\%$.
Besides, notes with the first interaction within 24 hours and 100 exposures increase by $9.83\%$.
\section{Related Work}
\label{rel}

\textbf{Multimodal Representation.}
With the advancement of pre-training techniques, numerous studies utilize web-scale image-text data to pre-train models for a fundamental understanding of multimodal information.
There are two main pre-training techniques.
The first technique adopts generative pre-training objectives, following masked language/image modeling~\cite{dou2022empirical,wang2023image,li2021align} or image captioning~\cite{yu2022coca,li2022blip}.
The second technique uses contrastive pre-training objectives, such as contrastive learning~\cite{radford2021learning,zhai2023sigmoid,sun2023eva,sun2024eva,yu2022coca,wang2023image} and image-text matching tasks~\cite{dou2022empirical,li2021align,li2022blip}.
These models demonstrate a powerful zero-shot ability in many downstream tasks.
However, the text comprehension capabilities of these models are unsatisfactory~\cite{chen2023difference,zhang2024long}, particularly in the era of LLMs.
In this work, we explore the use of LLMs' powerful comprehension capabilities to enhance multimodal representations.

\textbf{Large Language Models.}
LLMs have attracted extensive attention due to their remarkable text understanding and generation capability~\cite{jiang2023mistral,vicuna2023,bai2023qwen,touvron2023llama}.
Therefore, numerous text applications have been developed based on LLMs~\cite{ma2023fine,zhang2024notellm,muennighoff2024generative,xu2023large,karamcheti2024prismatic}.
To understand more modality information, LLMs are extended into MLLMs~\cite{bai2023qwenvl,chen2024tomgpt,yin2023survey,mckinzie2024mm1,laurenccon2024matters}.
Most MLLMs employ specialized modality encoders to capture non-textual information~\cite{su2023pandagpt,bai2023qwenvl,chen2024tomgpt}.
This information is mapped into textual space and processed by LLMs, enabling MLLMs to generate answers from multimodal input~\cite{ge2023planting,dong2023dreamllm,bai2023qwenvl,chen2024tomgpt}.
While existing studies primarily focus on the generation abilities of LLMs, some explore the text representation ability of LLMs~\cite{muennighoff2024generative,springer2024repetition,zhang2024notellm,jiang2023scaling,ma2023fine,behnamghader2024llm2vec}.
However, no studies have explored multimodal representation with LLMs. 
Therefore, we investigate the multimodal representation ability enhanced by LLMs in recommendation scenarios.

\textbf{Multimodal I2I Recommendation.}
I2I recommendation generates a ranked list of items from a vast pool based on item similarity, which is an important technique for recall stages in recommendation~\cite{yang2020large,zhu2018learning,zhang2024notellm}.
I2I recommendation employs a pre-constructed I2I index~\cite{yang2020large} to retrieve relevant items by the approximate k-nearest neighbor method~\cite{johnson2019billion}.
Some studies have investigated multimodal content-based I2I recommendations~\cite{yu2022commercemm,das2022maps,jin2023learning,zheng2023make,han2023fame}.
However, these works adopt small-scale pre-trained visual-linguistic models as backbones, which have huge potential for scaling up.
Therefore, in this work, we explore empowering the multimodal representations with LLMs in multimodal I2I recommendations.

\section{Conclusion and Future Work}
This paper explores using LLMs to improve textual comprehension in multimodal representation tasks under I2I recommendation scenarios.
We design an end-to-end fine-tuning method that can customize the integration of any existing LLMs and vision encoders.
This approach reduces reliance on open-source MLLMs, which require costly multimodal pre-training.
To prevent the problem of ignoring visual information in end-to-end fine-tuning, we propose \method, which contains mICL and late fusion.
The effectiveness of this approach is confirmed through extensive experiments.
In future work, we will extend our method to more complex modalities, like videos, and broader scenarios and recommendation tasks.

\section*{Acknowledgements}
This work was supported in part by the grants from National Science and Technology Major Project (No. 2023ZD0121104), and National Natural Science Foundation of China (No.62222213, 62072423). Xiangyu Zhao was partially supported by Research Impact Fund (No.R1015-23), APRC - CityU New Research Initiatives (No.9610565, Start-up Grant for New Faculty of CityU), CityU - HKIDS Early Career Research Grant (No.9360163), Hong Kong ITC Innovation and Technology Fund Midstream Research Programme for Universities Project (No.ITS/034/22MS), Hong Kong Environmental and Conservation Fund (No. 88/2022), SIRG - CityU Strategic Interdisciplinary Research Grant (No.7020046), Huawei (Huawei Innovation Research Program), Tencent (CCF-Tencent Open Fund, Tencent Rhino-Bird Focused Research Program), Ant Group (CCF-Ant Research Fund, Ant Group Research Fund), Alibaba (CCF-Alimama Tech Kangaroo Fund No. 2024002), CCF-BaiChuan-Ebtech Foundation Model Fund, Collaborative Research Fund No.C1043-24GF, and Kuaishou.

\clearpage
\balance
\bibliographystyle{ACM-Reference-Format}
\bibliography{sample-sigconf}

\balance
\appendix
\section*{Appendix}

\section{Training Details}
\label{para:Training Details}

\begin{table}[!h]
    \centering
    \setlength\tabcolsep{2mm}
    \renewcommand\arraystretch{0.9}
    \caption{Detailed training hyperparameters for fine-tuning MLRMs.}
\begin{tabular}{lc}
\Xhline{1.0pt}
\textbf{Configuration} & \textbf{Hyperparameter} \\
\Xhline{0.5pt}

Optimizer & AdamW \\
$\beta_1$ & $0.9$ \\
$\beta_2$ & $0.999$ \\
$epsilon$ & $1e^{-8}$ \\
Max gradient norm & $1.0$ \\
Weight decay & $1e^{-3}$ \\
Peak learning rate & $3e^{-6}$ \\
Warmup Ratio & $0.1$ \\
Learning rate scheduler & linear decay \\
Numerical precision & bf16 \\
Global batch size & $128$ \\
Training steps & $7,956$ \\

\Xhline{1.0pt}
\end{tabular}
\label{tab:traininghyper}
\end{table}

For all models with LLMs, we list the detailed training hyperparameter in Table~\ref{tab:traininghyper}.
Besides, we use DeepSpeed~\cite{rasley2020deepspeed}, and Zero Redundancy Optimizer (ZeRO)~\cite{ren2021zero} Stage 3 to train our models.

\section{End-to-End MLRM Details}
\label{para:MLRMs}
We provide the details about end-to-end MLRMs in Table~\ref{tab:modelhyper}.
Besides, we present the continuous pre-training details of Tomato to enhance understanding of its characteristics.
Tomato is based on LLaMA 2~\cite{touvron2023llama}, which originally only supports English. 
However, since our platform targets the Chinese market, it is necessary to enhance LLaMA 2's Chinese understanding capabilities. 
Thus, we have made the following improvements to continuous pre-training:

\begin{itemize}[leftmargin=*]
    \item \textbf{Vocabulary Expansion:} We extended the original vocabulary from $32,000$ tokens to $49,216$ tokens. The additional $17,216$ tokens are primarily Chinese.
    \item \textbf{Pre-training Data:} We collect pre-training data from three sources: Chinese corpora, English corpora, and Platform-specific Data.
     We utilize open-source Chinese corpora, including Wanjuan~\cite{he2023wanjuan} and WuDao~\cite{wudao}, totaling approximately 2T tokens.
     To maintain English proficiency, we also include RedPajama~\cite{together2023redpajama}.
     Besides, we incorporate 30B tokens of high-quality notes collected from our platform to better understand the data characteristics of our platform.
\end{itemize}
Tomato enhances Chinese understanding while maintaining its English capabilities at a level comparable to the original LLaMA 2 model.
It is worth noting that the pre-training details are unrelated to our contributions.
Our method is based on the model after continual pre-training.
This also shows an advantage of our method, which allows for the personalized enhancement of the base models for different modalities before fine-tuning using \method.

\begin{table}[!h]
    \centering
    \setlength\tabcolsep{1mm}
    \renewcommand\arraystretch{0.9}
    \caption{Model settings of end-to-end MLRMs.}
\scalebox{0.65}{
\begin{tabular}{lccc}
\Xhline{1.0pt}
\textbf{Configuration} & \textbf{MTomato-Base} & \textbf{MQwen-Base} & \textbf{MQwen-bigG} \\
\Xhline{0.5pt}
Vision encoder $V_{\theta}$ init. & CLIP ViT-B & CLIP ViT-B & OpenCLIP ViT-bigG \\
Connector $C_\pi$ init. & random Q-Former & random Q-Former & random Q-Former \\
LLM $LLM_\mu$ init. & Tomato & Qwen-Chat & Qwen-Chat \\
\Xhline{0.5pt}
\# Vision encoder layers & $12$ & $12$ & $48$ \\
\# Vision encoder attention heads & $12$ & $12$ & $16$ \\
Vision encoder sequence length $L$ & $196$ & $196$ &$256$ \\
Vision encoder hidden size $h_v$ & $768$ & $768$ & $1,664$ \\
Vision encoder intermediate size & $3,072$ & $3,072$ & $8,192$ \\
\Xhline{0.5pt}
\# Q-Former layers & $6$ & $6$ & $6$ \\
\# Q-Former attention heads & $12$ & $12$ & $12$ \\
Learnable query numbers $L_c$ & $16$ & $16$ &$16$ \\
Q-Former cross attention frequency & $2$ & $2$ & $2$ \\
Q-Former cross attention hidden size & $768$ & $768$ & $1,664$ \\
Q-Former hidden size & $768$ & $768$ & $1,536$ \\
Q-Former intermediate size & $3,072$ & $3,072$ & $3,072$ \\
\Xhline{0.5pt}
\# LLM layers & $32$ & $32$ & $32$ \\
\# LLM attention heads & $32$ & $32$ & $32$ \\
Vocab size & $49,216$ & $151,936$ & $151,936$ \\
LLM hidden size $h_t$ & $4,096$ & $4,096$ & $4,096$ \\
LLM intermediate size & $11,008$ & $22,016$ & $22,016$ \\
\Xhline{1.0pt}
\end{tabular}}
\label{tab:modelhyper}
\end{table}
\begin{table*}[!t]
    \centering
    
    \setlength\tabcolsep{8pt}
    \renewcommand\arraystretch{0.9}
    \caption{Zero-shot representation performance of different models under different modal inputs (\%). ``\textbf{{\Large *}}'' indicates the statistically significant improvements (i.e., two-sided t-test with $p<0.05$) over the best baseline.}
    \scalebox{0.85}{
    \begin{tabular}{l|l|ccc|ccc|ccc} 
    \Xhline{1.0pt}
    \multirow{2}*{\textbf{Method}}&\multirow{2}*{\textbf{Input}}&\multicolumn{3}{c|}{\textbf{All Pair}}&\multicolumn{3}{c|}{\textbf{Short Query Pair}}&\multicolumn{3}{c}{\textbf{Short Target Pair}} \\
    \cline{3-11}
    ~&~&R@100&R@1k&R@10k&R@100&R@1k&R@10k & R@100&R@1k&R@10k \\
    \Xhline{0.5pt}
    \multirow{3}*{BLIP-2}&Image&19.60&34.16&53.49 & 19.08 &34.92&56.18 &20.06 & 36.41 & 58.20 \\
    ~&Text&15.83&27.29&44.50&9.42&16.36&31.29&12.16&21.15&39.94 \\
    ~&Multimodal&27.28&43.00&62.69 & 22.46 & 36.93 & 59.54 & 25.16 & 40.54 & 60.92 \\
    \Xhline{0.5pt}
    \multirow{3}*{LLaVA-1.5}&Image&18.88&33.17&53.14&19.64&34.18&56.18&21.00&35.38&57.10 \\
    ~&Text&30.86&46.71&64.14&21.90&34.43&49.63&22.28&32.98&47.48 \\
    ~&Multimodal&37.01&55.45&73.72&30.50&47.78&\textbf{68.74*}&31.53&48.09&68.12 \\
    \Xhline{0.5pt}
    \multirow{3}*{Qwen-VL}&Image&17.55&31.62&51.58&17.77&33.55&55.56&18.41&33.88&55.97 \\
    ~&Text&38.74&54.65&70.37&23.31&35.17&51.95&22.17&34.84&51.29 \\
    ~&Multimodal&37.87&55.48&74.72&26.29&43.14&64.88&29.81&47.66&\textbf{69.68*} \\
    \Xhline{0.5pt}
    \multirow{3}*{Qwen-VL-Chat}&Image&17.43&31.15&50.48&17.59&32.29&55.04&18.75&33.40&55.09 \\
    ~&Text&37.92&53.02&68.48&22.63&33.87&49.15&21.32&32.50&48.03 \\
    ~&Multimodal&40.94&58.58&75.43&29.07&44.51&64.14&31.81&49.10&68.54 \\
    \Xhline{0.5pt}
    Qwen-Chat&Text&38.98&54.08&69.02&23.08&34.90&50.44&26.04&38.35&53.33 \\
    Tomato&Text&42.67&60.54&78.01&24.51&37.35&56.45&32.27&46.85&66.75 \\
    BM25&Text&\textbf{55.22*}&\textbf{72.64*}&\textbf{82.82*}&\textbf{35.75*}&\textbf{48.41*}&59.81&\textbf{35.05*}&\textbf{49.52*}&63.64\\
    \Xhline{1.0pt}
    \end{tabular}}
    \label{tab:zero-shot}
\end{table*}
\section{Saliency Scores of Qwen-VL-Chat}
\label{para:qwsaliency score}

\begin{figure}[!h]
    \centering
    \includegraphics[width=0.26\textwidth]{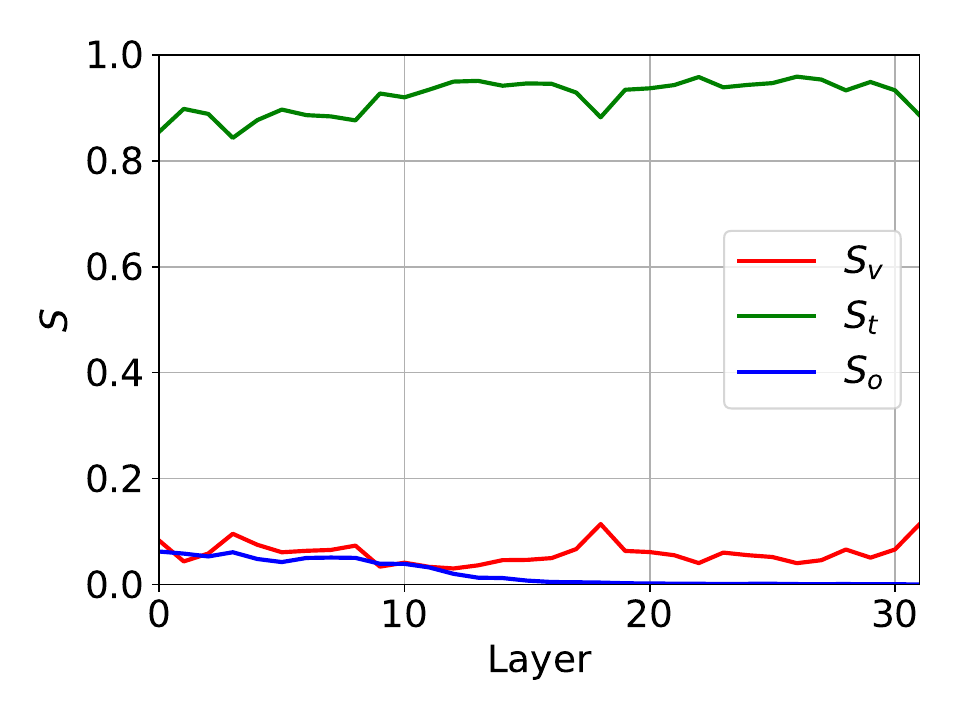}
    \caption{Relative sizes of $S_v$, $S_t$ and $S_o$ in different layers of Qwen-VL-Chat.}
    \label{fig:qw saliency score}
\end{figure}

We observe a unique pattern of saliency scores in Qwen-VL-Chat, as shown in Figure~\ref{fig:qw saliency score}.
The predominant information flow in Qwen-VL-Chat is textual, with a value higher than $0.8$ in all layers.
This result is attributed to the extensive visual embeddings ($256$ in Qwen-VL-Chat) inputted into LLMs, which reduces the average saliency of image information flow.
This denotes that traditional MLLMs are not directly suitable for MLRMs.
The necessary action to adapt MLLMs to MLRMs is to increase the density of information flow.

\section{Performance of Zero-shot Multimodal Representation}
\label{para:zero-shot}

In this section, we explore the multimodal representation ability of MLLMs in a zero-shot way in the multimodal I2I recommendation task.
We select four popular MLLMs, i.e., \textbf{BLIP-2}~\cite{li2023blip}, \textbf{LLaVA-1.5}~\cite{liu2023improved}, \textbf{Qwen-VL}~\cite{bai2023qwenvl} and \textbf{Qwen-VL-Chat}~\cite{bai2023qwenvl}.
We also compare some baselines \textbf{BM25}~\cite{robertson2009probabilistic}, \textbf{Qwen-Chat}~\cite{bai2023qwen} and our continual pre-trained LLM \textbf{Tomato}, which is pre-trained on our platform data based on LLaMA 2~\cite{touvron2023llama}, but lacks vision perception ability.
The MLLMs details are concluded in Table~\ref{tab:sum}.
To understand the representation ability of MLLMs for different modalities, we also input images and texts independently for the representation test.

The results are presented in Table~\ref{tab:zero-shot}.
We have several observations.
Firstly, the zero-shot representation performance of existing MLLMs is inferior to BM25.
This indicates that despite their excellent visual understanding, the zero-shot multimodal representation ability of MLLMs is unsatisfactory.
This is because MLLMs are trained using language modeling loss, which mismatches representation tasks.
Therefore, additional training is needed to properly align MLLMs with representation tasks.
Next, we find that MLLMs can achieve better performance for multimodal input in most cases, which demonstrates that MLLMs can extract and fuse multimodal information effectively.
At the same time, we discover that most MLLMs are better at representing text information than image information.
This is reasonable because most parameters of MLLMs originate from LLMs, and the text of notes is more discriminative than images in our scenarios.
Lastly, we find that LLMs can achieve performance comparable to MLLMs without any image input.

\section{Model Performance on Long Query and Target Pairs}

To comprehensively evaluate our method's effectiveness, we conduct tests on Long Query and Long Target Pairs.
Long notes are defined as those exceeding $165$ tokens (about $10\%$ of test notes).
The test dataset contains 2,228 long query pairs and 2,177 long target pairs.
The results are shown in Table~\ref{tab:long}.
Due to the excellent long-context understanding of LLMs, the performance on long pairs is significantly higher than on overall pairs, as long pairs provide more information. 
Additionally, our method improves upon the basic method for long pairs, which shows that enhancing the focus on visual information is also important for long pairs.

\begin{table}[!t]
    \centering
    
    \setlength\tabcolsep{2pt}
    \renewcommand\arraystretch{0.9}
    \caption{Performance of different methods based on three MLRMs on Long Query Pair and Long Target Pair (\%).}
    \scalebox{0.8}{
    \begin{tabular}{l|l|ccc|ccc} 
    \Xhline{1.0pt}
    \multirow{2}*{\textbf{Base Model}}&\multirow{2}*{\textbf{Method}}&\multicolumn{3}{c|}{\textbf{Long Query Pair}}&\multicolumn{3}{c}{\textbf{Long Target Pair}} \\
    \cline{3-8}
    ~&~&R@100&R@1k&R@10k&R@100&R@1k&R@10k \\
    \Xhline{0.5pt}
    \multirow{6}*{MTomato-Base}&basic method&80.41&95.06&98.93&78.21&94.07&98.56\\
    ~&mICL&80.70&95.50&99.37&79.64&94.12&98.81 \\
    ~&late fusion&80.85&\textbf{95.79}&\textbf{99.41}&79.84&94.36&99.01 \\
    ~&\method&\textbf{81.57}&95.55&99.32&79.34&\textbf{95.05}&\textbf{99.25} \\
    ~&only late fusion&80.80&95.30&99.32&\textbf{80.23}&94.76&98.91
 \\
    ~&Omni-Retrieval&79.98&94.68&98.98&77.77&93.13&98.67 \\
\Xhline{0.5pt}
\multirow{6}*{MQwen-Base}&basic method&80.80&95.02&99.23&79.99&94.47&98.86
\\
    ~& mICL&82.01&95.31&99.37&80.98&94.37&99.01
 \\
    ~&late fusion&\textbf{82.16}&95.50&99.42&81.07&94.51&99.11
 \\
 ~&\method & 82.01 &\textbf{95.99} & \textbf{99.47}&81.47&\textbf{95.01}&\textbf{99.31}\\
    ~&only late fusion&81.96&95.93&99.46&\textbf{82.11}&94.56&99.06 \\
    ~&Omni-Retrieval&80.95&95.55&99.23&79.25&94.32&98.72 \\
\Xhline{0.5pt}
    \multirow{6}*{MQwen-bigG}&basic method&82.78&96.17&99.60&82.31&95.75&99.01
\\
    ~&mICL&\textbf{83.36}&96.17&\textbf{99.61}&\textbf{82.51}&\textbf{95.80}&99.16
 \\
    ~&late fusion&82.65&95.93&99.56&82.26&95.16&99.21
 \\
 ~& \method&82.74&\textbf{96.32}&99.56&81.82&\textbf{95.80}&99.01 \\
    ~&only late fusion&82.64&95.79&99.46&80.78&95.35&\textbf{99.25}
 \\
 ~&Omni-Retrieval&82.30&96.03&99.41&80.63&95.16&99.11 \\
    \Xhline{1.0pt}
    \end{tabular}}
    \label{tab:long}
\end{table}

\end{document}